# Statistical Imaginaries, State Legitimacy: Grappling with the Arrangements Underpinning Quantification in the U.S. Census

**Jayshree Sarathy, Northeastern University, USA**[1]
**danah boyd, Cornell University, USA**[2]

**Abstract:** Over the last century, the adoption of novel scientific methods for conducting the U.S. census has been met with wide-ranging receptions. Some were quietly embraced, while others sparked decades-long controversies. What accounts for these differences? We argue that controversies emerge from and reconfigure *arrangements of statistical imaginaries*, putting into tension divergent visions of the census. To analyze these dynamics, we compare reactions to two methods designed to improve data accuracy—imputation and adjustment—and two methods designed to protect confidentiality—swapping and differential privacy, offering insight into how each method reconfigures stakeholder orientations and rhetorical claims. These cases allow us to reflect on how technocratic efforts to improve accuracy and confidentiality can strengthen—or erode—trust in data. Our analysis shows how the credibility of the Census Bureau and its data stem not just from empirical evaluations of quantification, but also from how statistical imaginaries are contested and stabilized.

## 1. Introduction

As the datafication of the state has expanded, more scholars have interrogated the function of government statistics (Mennicken and Espeland, 2019). But quantification is not a new phenomenon, and prior studies highlight how statistics are used in state-making efforts (Scott, 1999; Starr, 1987). From 'political arithmetic' in the seventeenth century (Desrosières, 2011; McCormick, 2009) to laws of large numbers in the nineteenth century (Hacking, 1990), governments have leveraged statistics to not only catalog and control their populations, but also to render themselves legitimate (Porter, 2020). The state relies on the authority of numbers (Power, 1997; Strathern, 2000) to assert the validity of its actions, resist politicization, and justify policies (Deringer, 2018; Murphy, 2017). But numbers are just as constructed as the administrative politics they support (Berman, 2022; Tooze, 2008). As the pressures on data continue to increase, we are seeing another side to the story; just as numbers can help create and maintain state legitimacy, they can also be used to destroy it (Didier, 2018; Rosa, 2014). In turn, crises of legitimacy can lead to further technical failures (Vertesi and boyd, 2023).

Nowhere is this balancing act more evident than in the project of population censuses, which are carefully-managed technical and bureaucratic operations (Anderson, 2015; Bryant and Dunn, 1995; Rodríguez, 2000). In the United States, scholars have been especially attentive to how measurements of race, ethnicity, and other demographic attributes are critical for operationalizing democracy yet leave the credibility of the state open to attack (Nobles, 2000). These debates over who is counted and how extend beyond experts. Most censuses are no longer just top-down, state-controlled operations, but ones that are actively negotiated by various social actors, including data users and community advocates (Emigh et al., 2016; Villacis et al., 2022) who often reject the epistemic commitments of technocratic actors (Eyal, 2019).

---

[1] The majority of this work was conducted when J.S. was at Columbia University.
[2] This work was conducted when d.b. was at Microsoft Research and Georgetown University.

Contestations around census data often focus on data quality, but this concept is more nuanced than it seems. Theories of quantification highlight how numbers are produced, when numbers matter, and what makes numbers credible (Berman and Hirschman, 2018; Hacking, 1990; Porter 2020; Power 1997). They often emphasize how the relationship between data quality and credibility is complicated; data deemed to be "low-quality" can unsettle the legitimacy of data projects (Ghosh, 2020; Sturge, 2022), even as "made up" data can be statistically constructive (Hacking, 2006; Lampland, 2010). Less attention has been paid, however, to the ways in which *entire* sociotechnical systems—including the state and its data—are made (un)stable or (il)legitimate in ways that are *disconnected* from the purported quality of the data.

In this work, we consider the relationship between quantification and legitimacy in the U.S. census beyond data quality. Through analyzing two pairs of case studies of methodological interventions, we explore the following questions.
1. What accounts for differing receptions to methodological advances? In particular, why do some advances trigger controversy while others do not?
2. How does the reception of modernization projects impact the legitimacy of the state's data infrastructure?

To unpack these questions, we draw on one conceptualization of the relationship between scientific projects and state legitimacy, which is the framework of *imaginaries*—collective visions of social order and desirable futures, co-constituted with media and technology (Jasanoff and Kim, 2009). The role of imagination has also been used to explore the formation of a collective identity for nation-states (Anderson, 1983; Taylor, 2004), the co-production of science and democracy (Ezrahi, 1990; Latour, 1993), risk and uncertainty (Beck et al., 1992), and movements of resistance (Roy, 2009). Imaginaries allow us to focus our attention on what diverse communities – or, in our case, invested census stakeholders – believe the technical project to be rather than focusing on what it is.

Imaginaries are more than a conceptual device; they also serve an infrastructural role, creating the foundation for the credibility of the state and its data (Parks, 2015). The downside of this is that threats to imaginaries can unsettle entire systems. For example, boyd and Sarathy (2022) demonstrate how the disclosure avoidance system used in the 2020 U.S. Census exposed fractures in the *statistical imaginaries* that upheld the constitutional vision underpinning the legitimacy of the census. Other studies show that pursuing higher quality data through modern statistical methods can threaten established narratives of census data and, thus, the credibility of the entire project (Alonso and Starr, 1987; Choldin, 1994). These narratives are not monolithic; a project as large as a national census is stabilized by *multiple imaginaries*—held by different communities, factions within communities, or even by the same individuals who cross communities of practice.

In this work, we move beyond the existing literature to explore how multiple imaginaries work together to (de)stabilize statistical projects in ways that do not directly follow from the quality of the data itself. We argue that the particular *arrangement* of disconnected but overlapping imaginaries—that is, the network structure and rhetorical (mis)alignments—matters for understanding both reception and legitimacy.

We focus on the U.S. census because its repeated methodological crises (Anderson and Fienberg, 2001; Persily, 2010) render these arrangements visible (Jackson, 2014; Star, 1999). As the census is a unique statistical undertaking, our analysis is attentive to its particularities while opening up ways of examining controversies in other contexts. While many operational and political issues surrounding the census have triggered controversies, those that emerge over statistical methodologies tend to be especially revealing of conflicting statistical imaginaries. We examine two pairs of cases where networks of people and rhetorical claims—along with dynamics surrounding contested methodologies—allow some methods to stabilize arrangements of census imaginaries and others to cause rifts (Barry, 2002; Pinch and Bijker, 1984).

Using these case studies, we highlight how reception to new methods depends heavily on how disruptive they are to the existing arrangements of imaginaries. In particular, we find that *visibility* and *complexity* of statistical methods shape how the methods will be perceived within the census ecology. When the Census Bureau adopts simple or easily explainable methods with limited visibility, controversies rarely emerge in the moment; instead, such adoption can solidify arrangements that pose challenges later on. On the other hand, highly visible and highly complex methods are not only ripe for controversy out of the gate, but can also trigger controversies about simpler, existing methods. We examine these dimensions of visibility and complexity to consider the role that statistical methods play in the stability of the arrangement of imaginaries. We conclude with reflections on making *legitimate* data while making *data* legitimate.

## 2. Background and Methodology

### 2.1 The Statistical Work of the U.S. Census

Census data, like other infrastructures (Parks, 2015; Star and Ruhleder, 1996), are both taken-for-granted and visibly contested. While many large and long-lasting data projects face modernization challenges, the political ramifications of the U.S. census repeatedly trigger controversies (Anderson, 2015) which are often on display through lawsuits and media spectacle. Since the early part of the 20$^{th}$ century, many of these controversies center on statistical methods (Bouk and boyd, 2021). While methodological battles also trigger controversies around other state-produced data—from weather (Daipha, 2015) to economic measures (Mitchell, 2008; Stapleford, 2009)—the struggles that the Census Bureau faces in producing, repairing, and protecting the legitimacy of its data directly reflect the political role of these data.

The U.S. census is political because census data configure politics. The project was designed by the Constitution to be the basis for apportioning political representation in Congress, and has become critical infrastructure for policymaking, funding allocation, research, and public use (Anderson, 2015). The Constitution demands an "actual enumeration" of the population (U.S. CONST. art. I, § 2, cl. 3), an idealized vision of a non-partisan, objective, and precise count. As the tensions between this vision and the practice of census-taking are notably fraught (Anderson and Fienberg, 2001; Choldin, 1994), lessons learned from census controversies can help illuminate how conflict might unfold in other statistical projects.

While the U.S. census project has been executed since 1790, it was institutionalized at the start of the twentieth century, when government reformers offered a vision for a strong public administration (Wilson, 1887) that leveraged science, rigor, and bureaucracy to ensure that much of government work remained neutral (Akin, 1977). The creation of the permanent Census Bureau in 1902 was an embodiment of this ideal. Civil servants within the bureau leveraged statistical methodology, innovative technology, and organizational bureaucracy to guard against explicit political interference while producing population data that has become fundamental to governance (Dupree, 1986; Schor, 2017). But debates over statistical methods are often entangled with deeper epistemic contestations over what it means to know, who has the right to know, and who defines what is known (Porter, 2020; Starr, 1987). In the process of advocating for new techniques—and opening up the black box of making data—the Census Bureau has dealt with increasing contestation at every turn (Akin, 1977).

The modern Census Bureau is a budget-constrained federal agency that has dual commitments: publishing high quality statistics and ensuring the confidentiality of its respondents. While the Constitution simply requires the census to produce a count of the residential population for reapportionment, a host of factors—including federal and state statutes, commitments to evidence-based policymaking, and judicial decisions—have led the modern Census Bureau to publish a wide range of statistical products and to make them available for the public (Lane, 2021). Meanwhile, the Census Bureau recognizes the importance of confidentiality to ensure participation in the census, and laws and norms have solidified the bureau's obligations and moral commitments to protect data subjects. These dual commitments to publication and privacy are entangled and have been on a collision course for decades (Bryant and Dunn, 1995).

The increased use of census data for policymaking, politicking, and resource allocation has made them more entangled with the credibility of the state. In other words, quantification mediates an increasingly expansive yet fragile regime (Barry, 2002). As we heard from our interviewees, the categories of data that have been released in prior iterations of the census are now expected to be released in perpetuity with the same methods and better quality. At the same time, old methods must grapple with new critiques and better—but more complex—alternatives.

Thus, modernizing the census is no straightforward task. Methodological advances at the Census Bureau have often been met with widely different receptions (Cantwell et al., 2005; Rosenthal, 2000). Some developments are quietly accepted; others trigger existential threats. What makes a particular modernization effort controversial is as much about the methodology used as it is about the sociotechnical context in which that transformation unfolds.

### 2.2 Data and Analysis

In this paper, we analyze four different methodological advances to the U.S. Census: imputation, swapping, adjustment, and differential privacy. We draw our conclusions from interviews, observations, and archival records. The second author conducted four years of ethnographic fieldwork (2018-2022) within the U.S. Census Bureau and among external census stakeholders, focusing on broader questions of what makes data legitimate. This included over 1000 hours of direct observation of team meetings and public events in addition to hundreds of informal with civil servants, political appointees, and other census stakeholders. To go deeper on certain issues, she conducted semi-structured interviews with 83 people who worked on the 2020 census.

The first author has been involved with organizations from 2018-2024 that bring together technical stakeholders around modernizing privacy protections. As a computer scientist engaging with social scientists and government employees, she has observed first-hand the debates around developing new statistical tools for confidentiality in the 2020 census. She has additionally conducted 46 semi-structured interviews with stakeholders invested in access to federal statistics, including data archive administrators, developers of data analysis software, and researchers.

After comparing interview findings and field observations, both authors conducted archival work on the four controversies that were repeatedly highlighted in individual conversations. We collected and analyzed documents from census reports and scientific research papers, as well as arguments made during public events, on social media and blogs, and in court filings and news media. We followed up with existing informants to better understand what took place. We specifically focused on the motivation for and reception towards each methodological change. Overall, we do not aim to provide a perfectly comparative analysis across all four controversies but rather contribute a reflexive account of these changes to the census, situated as we are while one controversy is still roiling the census.

Our analysis process followed a reflexive thematic approach (Clarke and Braun, 2017), where we drew out themes in our data via memos, collaborative diagramming, and numerous discussions with each other and our informants. We iteratively mapped networks, agents, and attitudes to make sense of each case study, finding that the arrangements of perspectives were more informative than particular actors or agendas. We aim to pull out dynamics in a descriptive, rather than predictive, manner. This paper relies on these diverse data sources and analyses to develop a theoretical perspective on data legitimacy in periods of methodological change.

The census is a unique statistical undertaking. However, most statistical projects face localized controversies and we believe that unpacking the upheavals of one project—in this case, the US census—can open up ways of examining the controversies in another.

## 3. Arrangements of statistical imaginaries

Statistical imaginaries are infrastructural visions of institutions' knowledge production, upheld by both networks of people and narratives about data. Unlike sociotechnical imaginaries (Jasanoff and Kim, 2009), statistical imaginaries focus on the relationship between institutional power and statistics, encompassing not only desirable futures but also idealized pasts. Regarding knowledge about populations, statistical imaginaries can play a role in shaping and upholding demographic and state imaginaries (Brissette, 2016; Joyce and Mukerji, 2017; Rodríguez-Muñiz, 2021), but statistical imaginaries refer more specifically to conceptualizations of how data is collected, how data is processed via statistical techniques, and to what ends.

The prevailing statistical imaginary of the U.S. census may appear monolithic, but in practice, we found multiple—sometimes coherent, often contradictory—visions of what the census is or should be held by different stakeholders, including communities of practice, data subjects, and government officials. Disagreements over *who* to count, *how* to count them, and *where* they should be counted have repeatedly triggered controversies (Anderson and Fienberg, 2001; Persily, 2010). These controversies make visible the tensions between different imaginaries and their impact on the credibility of the project.

When using controversies as our entry point, we found that the lines between contestation, spectacle, and controversy can be blurry. We looked to the emergence of lawsuits to trace when and how routine debates turn into full-blown controversy. While legal challenges make visible the existence of a controversy, they also obfuscate the dynamics of the debate. Complex arrangements are forced into the simple binary of plaintiff and defendant, dividing networks of experts and advocacy groups into two sides and providing a compelling story of conflict for the media. Behind the scenes, the alignment of stakeholders and statistical imaginaries is often much more nuanced. Thus, we look to technical and organizational positioning, as well as legal stances, to make sense of each controversy.

### 3.1 Unpacking arrangements

In trying to understand what creates controversy, we find that how statistical imaginaries are *arranged* matters. Arrangements of statistical imaginaries speak to the configuration of overlapping and entangled visions that shapes both the work of the Census Bureau and the legitimacy of its data products. Even when different stakeholder groups have different imaginaries, they may end up supporting the same statistical methods. Yet, disrupting a given imaginary may cause enough dissent from the groups that held it, and potentially even destabilize the surrounding imaginaries, to threaten the entire arrangement. In other words, because there are divergent imaginaries at play, it is important to examine how seemingly disjointed imaginaries are stabilized in relation to one another through people and rhetoric.

A given arrangement is comprised of multiple imaginaries, which we observed via networks of stakeholders and sets of rhetorical claims, as described below. The stability of the arrangement, then, relies on the ways in which these networks and narratives intersect to uphold each other.

> *Stakeholder networks.* Any sociotechnical system is supported by communities of practice and networks of invested people (Star and Ruhleder, 1996), also known as stakeholders. Stakeholders can stabilize or upend a sociotechnical arrangement (Davis, 2023; Epstein, 1996). In our context, legitimacy is entangled with the structure of these networks as self-identified "census nerds" organize themselves to support or challenge the work of the government. Many stakeholders—from government officials to politicians to civil society organizations—play a role in upholding these overlapping visions. The structure of stakeholder networks is never stable – individual stakeholders are sometimes in coalition and in other cases in opposition with one another. Given the political nature of this project, many are invested in shifting the network structure itself. Moves to build or break parts of the network affect the legitimacy of the census project.

*Rhetorical claims.* Rhetoric is used to configure how different stakeholders understand sociotechnical systems. The management of competing rhetorical claims, and the resulting interpretative flexibility, shapes sociotechnical visions (Pinch and Bijker, 1984). Through rhetorical moves that build shared understanding or collective illusions, stakeholders can come together. Likewise, rhetoric can be used to divide stakeholder networks. Stable arrangements arise from achieving not just technical closure, but also rhetorical closure that knits together idealized visions of pasts, presents, and futures.

We analyze arrangements at the level of networks and rhetorical claims, for these are the most visible through our data. Through mapping actors involved in each controversy, we consistently saw that neither "role" nor "group" provided a viable orientation for analysis. Individuals within groups could have distinct imaginaries; sometimes, the same individuals (usually those in translational roles across communities) held multiple imaginaries depending on their interactions.

### 3.2 Controversy around methods

In the case of the census, stakeholder networks and rhetorical claims are configured in relation to statistical methods. Controversies around the census often appear to center statistical methods, but they are rarely about the methods themselves (Bouk and boyd, 2021). Frequently, controversies emerge when stakeholders have a vested interest in shaping the practices and outcomes of the census for specific purposes. For example, statisticians may expose statistical shortcomings of census data to improve downstream inference, while advocacy groups or politicians may be motivated by ideological goals or differential impacts. In shaping the data towards certain goals, these stakeholders seek to center a particular statistical imaginary, or to influence the entire network (Castells, 2011). Debates about methods, then, become fodder for battles over the arrangement of statistical imaginaries and the power of these arrangements. In our interviews, we found that some stakeholders firmly believe that the issue is methodological while others acknowledged the political agendas that shaped their stance.

Of course, not all disagreements about method erupt into a controversy (e.g., counting children in boarding schools (National Research Council, 2006)) and not all controversies are driven by statistical method (e.g., the citizenship question (Levitt, 2019)). Indeed, many controversies that surround the census—such as which questions to ask (Mora, 2014), who should be counted where, and when the count should take place (Anderson, 2015)—are well-explained through partisan politics. What makes methods-based controversies more interesting for analysis is that stakeholders cannot be easily clustered by political party. This does not mean that political calculations do not play a role, but that partisan positioning alone cannot explain the eruptions that take place.

While controversies around method can be hard to trace from beginning to end, we are better able to see different imaginaries and the ways in which the arrangement can amplify or stifle controversy. For example, technical evaluations by statisticians can become fodder for advocacy groups seeking to identify data quality concerns. These concerns can play into the agendas of partisan groups that aim to delegitimize the data as a whole. Media coverage can ramp up into court cases that constrain the technical leeway of the data production. Thus, a technical intervention aiming to improve data quality can end up imposing constraints that make data quality worse, impeding low-stakes communications with stakeholders who would otherwise engage in productive conversations. Grappling with the nature of these entanglements—and how they are shifted by the affordances of new methods—is essential for understanding these controversies.

### 3.3 What shapes the reception of methodological interventions?

In unpacking what helps explain differences in reception between different statistical methods, we find that methods that are *complex* and *visible* create challenges for achieving rhetorical closure. Complexity accounts for the level of specialized knowledge needed to understand a certain method, and the difficulty around explaining the method to a wide variety of stakeholders. Visibility denotes the breadth and magnitude of awareness of a particular method to different stakeholders. Sometimes, the Census Bureau creates visibility (beyond just transparency of methodological details) to elicit engagement from stakeholders; in other cases, stakeholders increase the visibility to put pressure on the bureau.

In analyzing our cases, we argue that statistical methods that afford low visibility and low complexity are less likely to create conditions that might destabilize the network of stakeholders and, thus, the arrangement of statistical imaginaries. Conversely, actors are better poised to leverage complexity or heightened visibility to challenge the legitimacy of the system.

Both dimensions are important. For example, visibility on its own does not always ensure controversy. There are several methods used in the census that are visible but not overwhelmingly controversial, such as the method of self-response using paper forms; these methods tend to be simple and rhetorically hard to contest. Similarly, complexity on its own does not imply visibility; there are methods that are operationally complex yet hidden from view, such as processes for age smoothing (i.e. adjusting responses for age that are rounded to the nearest 5 or 10). Complex operations tend to become visible as challenges to the procedure emerge. Complexity and visibility are thus neither independent nor fully aligned.

In choosing to examine visibility and complexity, we are not arguing that these are the only dimensions to consider—nor that these have explanatory power. Other methods-based controversies might require examining different dimensions. Rather, by teasing out these dynamics, we can start to see our cases in a new light and develop levers by which to unpack entangled imaginaries.

### 3.4 Productive destabilizations

Though unstable arrangements can cause controversy, stability should not be the only goal of statistical projects. Instability can be productive in breaking down brittle arrangements and opening up space for new ones to form. Methods that generate complexity and visibility may cause rifts, yet ultimately shift arrangements in ways that enable further improvements to data infrastructures. Thus, our analysis does not imply that organizations should reduce transparency and curb modernization efforts, but rather they should be aware of possible shifts so that they can work towards building more resilient, long-lasting data infrastructures.

## 4. Navigating a presumption of imperfect data

To appreciate how controversies rely on and reconfigure certain arrangements of statistical imaginaries, we turn to a pair of controversies involving four distinct methodologies used in the US census. The first pair of methods—imputation and adjustment—were designed to improve the quality of the data. The second pair of methods—swapping and differential privacy—were developed to limit the disclosure of sensitive information about individuals through statistical products.

*Pair #1: Methods to improve data accuracy*

Missing data has always been a problem for the census. There are missing addresses, missing households, missing individuals within households, and missing characteristics about individuals—all for various reasons. Missing data is not evenly distributed across the population. Certain populations are more likely to be missing in the data; these are referred to as *differential undercounts* (O'Hare, 2019).

Those who work on the census have repeatedly tried to address missing data through improvements to operational processes, including increasing the number of knocks on doors, hiring enumerators who speak different languages, and creating multiple pathways to respond (Prewitt, 2016; Rodríguez-Muñiz, 2021). But when self-response and fieldwork fail, those conducting the census use alternative methods to increase completeness.

Imputation: In the 1940s, the Census Bureau began experimenting with *imputation,* a "repair" technique developed to fill in whole households—and the characteristics of the people who are missing (Scheuren, 2005). Today, there are two types of imputation: *characteristic imputation*, which is used when a person exists but is missing details such as age, sex, or race, and *count imputation*, also called *whole person imputation*, wherein the presence and characteristics of an individual must be produced (Cantwell, 2021).

The earliest approaches to imputation involved filling in the existence of missing people using a random set of values on a punch card. In 1960, the census moved onto "hot-deck" imputation, where the recent, "hot" punch cards from similar households in the same census tract are copied and pasted into missing slots (Andridge and Little, 2010). Hot-deck imputation is consistent with an understanding of the census as a manual, enumerative (as opposed to statistical) operation, even though as we discuss in the next section, not all statisticians hold this vision. This imputation technique also presumes that the data is reasonably homogenous and thus, is rarely used in modern applications.

Adjustment: Imputation addresses missing data, but it does not solve differential undercounts. After decades of failing to remedy this systemic issue through operational advancements, the bureau started exploring statistical approaches for use in the data processing phase of its work.

Based on the success of its imputation work in the 1970s, the bureau explored a technique known as *statistical adjustment*.

Adjustment attempts to resolve biases within the data by replacing traditional census counts with statistically adjusted population figures (Choldin, 1994; Rosenthal, 2000). Adjustment was first proposed by bureau statisticians in advance of the 1980 census, but internal statisticians concluded that it was not sufficiently validated. The bureau revisited this approach in 1990, triggering a series of lawsuits (Department of Commerce v. United States House of Representatives, 1998; Wisconsin v. City of New York, 1996). In the end, the Supreme Court ruled that the Census Bureau and the Department of Commerce had the right to decide whether to use adjustment for redistricting data. After seeking guidance from civil servants at the Census Bureau, where there was not consensus, Secretary of Commerce Mosbacher decided not to go forth with the plan (Anderson and Fienberg, 1999). Those in favor of adjustment believe that this decision was politically motivated, while others defend this decision on technical grounds (Freedman, 1993).

In 2000, the Census Bureau proposed to use statistical sampling to follow up with non-response (U.S. Census Bureau, 2009), but this plan was abandoned after a Supreme Court ruling prohibiting the use of statistical sampling for apportionment data (Cantwell, 2021). By this point, Congressional attitudes towards adjustment were divided along partisan lines. The Census Bureau opted to cease investing in adjustment techniques (Choldin, 1994).

The so-called "adjustment wars" triggered fragmentation within critical networks of people. Inside the Census Bureau, factions formed and long-term civil servants quit in response. Academics took sides in the courtroom and in journals, splitting the field of statistics. Quickly, a wider array of stakeholders began questioning the validity of the data. Through litigation and Congressional hearings, adjustment became a more entrenched political issue. While government methodologists continued to investigate adjustment for the 2010 census, they opted—due to the politicization of the issue—to not even build the tools necessary to evaluate the feasibility of adjustment for the 2020 census.

<u>Adjustment impacting imputation</u>: The "adjustment wars" also sullied the reputation of imputation. Even though forms of imputation had been used since the 1940s—and the "hot deck" technique since 1960—those who got involved in adjustment litigation began calling imputation into question as well. This led to Utah arguing in 2002 that imputations of people in housing units for which the household size was unknown constituted an illegal use of sampling (Utah v. Evans, 2002). The Supreme Court ended up sustaining the use of hot-deck imputation, ruling that imputation differed from sampling in terms of "nature of the enterprise," "methodology," and "immediate objective" (Cantwell et al. 2005). Notably, the Supreme Court accepted the Census Bureau's argument that not imputing is equivalent of imputing zero and thereby intentionally creating an undercount (Hogan, 2001).

While the Supreme Court ruled in favor of the Census Bureau, the ruling triggered anxiety and conservativism within the bureau. Fearing another litigation fight, the Census Bureau opted not to leverage scientific advances in statistical methods to modernize its whole person imputation method, unsure if a new approach by the same name would also be supported. As a result, even though the scientific community has developed improved imputation techniques to account for heterogeneity in data, the Census Bureau continues to use hot-deck imputation because it is viewed as legally permissible (Cantwell, 2021).

*Pair #2: Methods to ensure statistical confidentiality*
Prior to 1840, the results of the census were posted publicly in the town square, with the goal of ensuring that everyone trusted the count (U.S. Census Bureau, 2019). But after the 1840 census, experts working for the government started to recognize that, without a commitment to confidentiality, the public would refuse to contribute their data. To remedy this, the Census Office of 1850 decided to end the practice of public validation and commit to the confidentiality of individual responses. This rhetorical commitment has persisted ever since, even with the notable failure to abide by said commitment in 1942 (Seltzer and Anderson, 2001, 2007).

The commitment to confidentiality was enshrined into law in the twentieth century, first under individual Census Acts and then permanently with Section 9 of Title 13 of the U.S. Code, which prohibits the release of information that could enable individual data subjects to be identified.

Throughout most of its history, the Census Bureau has guaranteed confidentiality by simply not publishing data. But during the 20th century, the bureau faced increased pressure to publish more data at more granular levels. Starting in 1960, the Census Bureau began strategically *suppressing* the publication of data that would be especially revealing (e.g. cells within tables that contained small counts) while also increasing the number of statistical tables it did publish. Recognizing the pending collision course between data access and privacy, the bureau also began researching new methods for disclosure avoidance (U.S. Census Bureau, 2019).

Swapping: In 1990, after deciding to publish more data at a granular level, the Census Bureau shifted from suppressing some data to strategically introducing noise via *swapping* (Fienberg and McIntyre, 2005), which involves taking households from two distinct locations and switching them. The goal is to ensure that outlier households are not easily identifiable from the published tables. Although the bureau also experimented with other techniques, swapping became the dominant technique used in 2000 and 2010.

Even though swapping introduced noise into the data, it was not seen by data users as controversial or threatening to the enumeration. Like imputation, it quietly made its way into standard census-making procedures. When data users learned about its usage, they were able to easily understand it.

Differential privacy: While swapping initially appeared to balance the dual commitments of confidentiality and data availability, computer scientists showed that it was possible to reverse engineer statistical tables into individual records (an act known as *reconstruction*) and then link those records to other datasets to *reidentify* individual respondents (Dinur and Nissim, 2003; Dwork et al., 2017). These findings called into question the disclosure-avoidance procedures that the bureau had been using since the 1990s. An internal investigation conducted in 2016 showed that such techniques could make tens of millions of individual records vulnerable to disclosure (Abowd and Hawes, 2023).

In 2018, the Census Bureau announced its intention to modernize its procedures for the 2020 census (Abowd, 2018b), arguing that differential privacy was a more principled way to satisfy its competing commitments (Abowd and Velkoff, 2019). Differential privacy is a mathematical framework that involves strategically injecting noise into statistical releases to limit individual-level disclosures (Dwork et al., 2006). For many stakeholders, however, differential privacy seemed like a dramatic shift from previous approaches. Some stakeholders who attended events sponsored by the Census Bureau were unaware that the bureau had been introducing noise into its published data since 1990, and the new approach unsettled their understanding of how census data are produced.

The announcement occurred in the midst of intense politicization of the 2020 census, prompting a range of stakeholders to interpret the new disclosure avoidance system as an existential threat (Ruggles et al., 2019). Many were concerned about the ways in which data quality would be impacted (Hotz and Salvo, 2020). Some data users also felt that the effort was unnecessary, given that there have been no external attacks on census data. The Census Bureau defended its move by pointing to the confidentiality mandate within Title 13 (Hawes, 2019: 13), yet critics challenged the bureau's interpretation of the law.

Iterative attempts to engage stakeholders backfired, undermining goodwill and fragmenting the network of stakeholders, largely because of the mismatched imaginaries at play (boyd and Sarathy, 2022). The controversy reached a boiling point when the state of Alabama sued the Census Bureau (Alabama v. United States Department of Commerce, 2021). Although the case was dismissed due to lack of standing, the unrest continued to brew.

Even with heightened controversy and another set of lawsuits, the Census Bureau went forward with its plans to implement differential privacy. Unlike with adjustment, the method made its way into the released data products (Abowd and Hawes, 2023). But the fights that emerged during the 2020 census are both over and not. Whether this method is accepted as legitimate depends on the arrangements of statistical imaginaries of those invested in federal data.

## 5. Analysis: The Structural Facets of Controversies

This bring us to our first research question: what shapes the reception of data modernization projects? In particular, why do some of these data modernization projects create more controversy than others? We find that when new methods are introduced, their reception is shaped heavily by how disruptive the methods are to the arrangement of imaginaries. Specifically, statistical methods that afford low visibility and low complexity (e.g., imputation and swapping) are likely to pose little disruption to the arrangement because they do not disrupt the network of stakeholders or upend existing rhetorical closure. But these dimensions may not remain fixed; a less complex method is often a precursor to more complex methods that aim to provide higher quality data but can wreak havoc on the arrangement.

Complex methods, on the other hand, become especially ripe for controversy once they are made visible (e.g., differential privacy). Heightened visibility (e.g., adjustment) can also trigger controversies around simpler systems (e.g., imputation). Moreover, antiquated methods (e.g., hot-deck imputation) can get locked into place once a controversy has been triggered, creating even greater hurdles for future campaigns around data improvement.

Figure 1 illustrates the four case studies we analyze in terms of their complexity, visibility, and level of controversy. The variation across these case studies shows how complexity and visibility are separate yet can interact in nuanced ways. Both dimensions are important for understanding how methods align with arrangements of imaginaries and potentially trigger controversies.

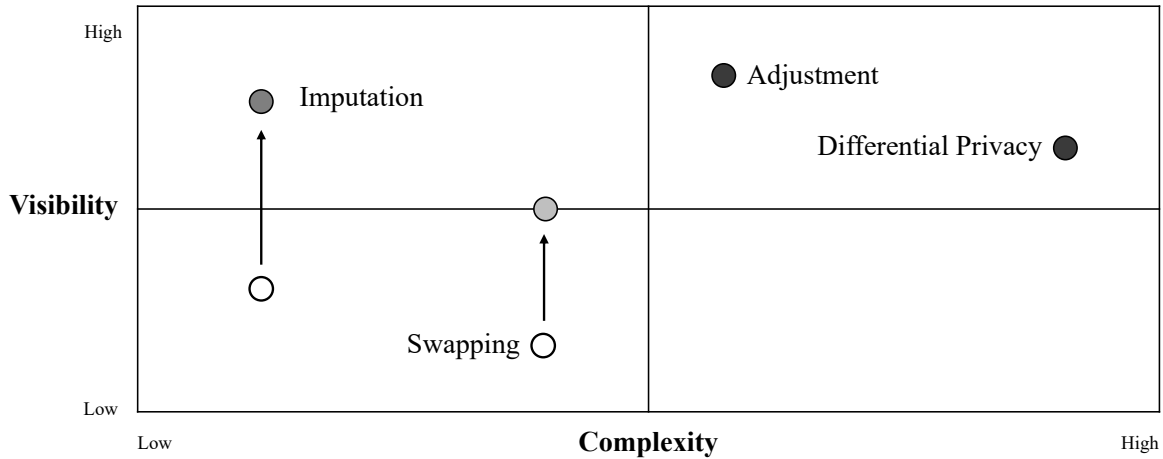

*Figure 1: Summary of case studies in terms of complexity and visibility. The shading of each circle indicates level of controversy, and the arrows indicate changes over decadal counts. As the case studies are nuanced, the figure is a simplified illustration rather than a full picture of the dynamics at play.*

Neither complexity nor visibility are inherent properties of any given methodology nor predictors of controversy. They must, instead, be understood in light of the networks of people and rhetoric at play. The larger and more diverse the network of stakeholders are, the more likely they are to find any given statistical method confounding. This is especially true when the rhetoric around the method signals that it is complex. After all, plenty of seemingly simple methods may be relatively easy to explain in ways that belie the impact they have on the data. The visibility of a method is also not a permanent state. As we can see with imputation, this was a method with low visibility until controversies over adjustment prompted attention to imputation.

We will examine each pair of case studies to explore these dynamics in more detail.

**5.1** *Imputation and swapping: reinforcing statistical illusions, at first*

When they were first introduced, imputation and swapping did not noticeably disrupt the arrangements of imaginaries. This was in large part due to their low visibility and low complexity. With both imputation and swapping, the Census Bureau did not hide its practices, but it did not publicize these techniques either. Implementation details were buried in technical reports, footnotes, and dry esoteric documentation. While the bureau was clear that its data processing work was designed to improve the quality of the data and ensure confidentiality, it did not call attention to the methodological details; there was no media coverage nor public requests for comment. Even the Census Bureau's official scientific advisory board paid little attention to these methods.

When asked, most statistical experts who were aware of this work did not believe that either imputation or swapping had much effect on total survey error. For its part, the Census Bureau never addressed either method in error estimates that it published for accountability purposes. Government publications led stakeholders to believe that these methods were applied on a small-scale, which added to their lack of visibility. For example, the bureau highlighted how imputation has been applied to less than one percent of data records in each decadal count (Cantwell, 2021). The Census Bureau has never publicly stated the rate of swapping, often invoking the notion of "security through obscurity," except to say that this mechanism did not materially change the on-the-ground enumeration (U.S. Census Bureau, 2021).

Even once they became more visible, the methods used for imputation and swapping are relatively simple to explain. Hot-deck imputation can be described as copy-and-pasting records from similar households, and swapping can be explained as switching the outlier households within geographic regions. Most stakeholders we spoke with understood the concepts, even if they lacked appreciation for the downstream implications of such methods. The simplicity of these narratives is critical for external communication as well as internal consensus within the bureau. Importantly, the explanations of both methods align with imaginaries of the census as strict enumerations, even if the impact of these methods is more nuanced.

While explaining these methods is relatively easy, explaining their statistical impact is not. Statistical experts who understood the ramifications of these methods had grown uneasy about their use because of their deeper understanding of the effects on the statistical work (National Research Council, 2009). However, simplicity had meant that the rhetorical claims around imputation and swapping did not initially destabilize the networks of data users and experts.

In addition to being straightforward to explain, these methods were also simple to implement. Both methods were operationally narrow in that they were each implemented by a small team within the Census Bureau at discrete points in the data production pipeline. Because they altered the numbers in a way that did not upend organizational processes, they were perceived as non-complex also within the census infrastructure.

Because of limited visibility and limited complexity, imputation and swapping did not initially disrupt the arrangement of statistical imaginaries. Many statisticians supported the use of these statistical techniques, which at the time were cutting-edge, as imputation would improve data quality and swapping would enable the release of granular data. Widespread imaginaries of strict enumeration (held by some groups of social scientists, redistricters, lawmakers, advocacy organizations, and the public) were also able to remain intact: both methods are compatible with commonly held statistical illusions, such as treating the data as accurate down to the individual level without the need for error bars (boyd and Sarathy, 2022), or by viewing outliers as the only data points vulnerable to re-identification attacks. And since the rhetoric of how the data should be used and perceived did not change, the network of stakeholders continued to work together to uphold the credibility of the data. Their concerns about data quality diverged, but neither saw these techniques as central to their concerns.

However, as statisticians' modeling techniques became more advanced, it became increasingly clear to technical communities that these simple methods of imputation and swapping were built on faulty assumptions. For example, hot deck imputation was predicated on the idea that neighbors were similar to one another, which was increasingly untrue (National Research Council, 2009). Likewise, swapping was predicated on the idea that increasing the homogeneity of a specific geography would have limited statistical implications; this was becoming increasingly untrue as the need for heightened swapping rates increased (Christ et al., 2022). The fact that error rates increased with the use of the method rankled the sensibilities of many statisticians. Inferential statisticians within and outside the bureau could no longer unequivocally advocate for swapping and imputation as they had been implemented within the census (Abowd, 2018a; Rubin, 2018). It is these expert actors who pushed hardest to be more transparent and leverage more complex approaches that produced less statistical error. In the process, they helped trigger controversies around the very systems they were dedicated to improving.

**5.2** *Adjustment and differential privacy: triggering shifts to the arrangement*

Visions of *how* to achieve "better" data are not shared by all stakeholders. A large community of inferential statisticians believes that adjusting population counts creates more accurate data *as a whole* and that differential privacy ensures better *overall* statistics while ensuring confidentiality. But not everyone agrees that these modernizations are warranted or beneficial. To understand what made adjustment and differential privacy so controversial for a range of census stakeholders, we once again return to the viewing these methods not in terms of their supporters and critics, but in terms of the larger arrangements of imaginaries at play in each case.

By the middle of the 20th century, discussions of population undercounts were widespread. Statistical adjustment was regularly floated as a remedy. In 1978, the National Academies of Science, Engineering, and Medicine published a panel report that argued that, regardless of what the Census Bureau said, statistical adjustment was technically feasible (National Research Council, 1978). Thus, even before the Census Bureau began conducting evaluations and tests for the 1980 census, adjustment was highly visible to a range of stakeholders. While initially non-controversial as a research endeavor, this approach became visible to more stakeholders through a series of court cases (Choldin, 1994).

The complexity of adjustment stemmed from its scale of impact. Unlike methods like imputation, adjustment was designed to touch *all* data to improve the macro statistical view. This aligns with a view, held predominantly by inferential statisticians, that the census is a continually evolving statistical project that is liable to systemic biases seen in other forms of surveys, and that producing a responsible census means correcting these biases with the help of other data sources. However, this deviated from others' belief that the census was nothing more than an enumeration of the total population. For researchers in various local and federal government agencies, adjustment flew in the face of their view of the census as a standalone data project that should not be influenced by other data sources, and as a politically precarious endeavor that should remain free from large-scale manipulations—even if these manipulations would reduce total survey error. The Census Bureau struggled to explain how altering local counts improved the fairness of the overall count, adding to the sense of complexity and creating challenges for rhetorical closure.

As with adjustment, differential privacy had a widespread impact on the data. In addition, it is even more technically complex to understand and implement. Informally, the 2020 implementation of this method adds noise to population counts at different geographic levels with an eye to maximizing statistical accuracy while minimizing risks of individual disclosure. As an approach that ensures the macro view of data remains consistent, it is fundamentally at odds with the idea of micro-level swapping, whose approach of minimizing edits to individual records compromises the macro view (Abowd, 2018a). As with adjustment, the idea that population-level alterations would increase the quality of the data overall challenged the prevailing statistical imaginary that population-level data is nothing more than the sum of individual records.

To its supporters, including computer scientists, economists, and government statisticians, differential privacy was aligned with their statistical imaginary. Moreover, this approach satisfied their beliefs that privacy is a core commitment that should be rigorously maintained, that noise should be added to data transparently so that it can be incorporated into downstream inferences, and that such noise should be carefully calibrated to optimize the high-level accuracy of the data.

Unfortunately, advocates of differential privacy could not give clear rhetorical support to the Census Bureau's implementation. While differential privacy is already complex, its deployment was additionally nuanced, as government statisticians had to innovate to make it work within the operational and statistical constraints of the census (JASON, 2020). Some technical stakeholders felt as though the bureau had undermined its responsibility to confidentiality by relaxing its approach in order to match existing systems and appease critics.

Like adjustment, the intention to implement differential privacy was made visible even before the Census Bureau was able to finish implementation and conduct evaluations. This was done intentionally by the Census Bureau for the sake of transparency and stakeholder engagement. Almost immediately after deciding to pursue differential privacy, the Census Bureau began discussing its plans publicly and inviting public feedback (Abowd, 2018a). Bureau leadership did not realize that their efforts to be transparent would reveal divergences in the imaginaries of stakeholders. By engaging in conversations around disclosure avoidance, stakeholders began to understand that noise infusion had already been happening at significant levels since 1990 through methods like swapping. As such, differential privacy also lifted the curtain around disclosure avoidance in the census as a whole. This increase in visibility was not paired with rhetoric that was interpretable to observers. To the contrary, efforts to explain differential privacy to the diverse range of stakeholders who invested in the plans failed to reassure them. This *legibility* crisis only increased the *legitimacy* crisis (Abdu et al., 2024).

Both adjustment and differential privacy were much more complex than their predecessors. Rhetorically, the narratives around adjustment and differential privacy were sufficiently complex that their implementations were not well-agreed upon (Freedman, 1993; JASON, 2020). This might have been true for swapping and imputation as well, but the lack of visibility combined with the low levels of complexity of these methods created a perception of simplicity and reliability once they came into focus, which was long after implementation and only after controversies erupted around differential privacy and adjustment. Because both adjustment and differential privacy were visible before implementation and viewed as complex and even experimental, those working on these implementations were not able to create consensus either internally or externally in advance of deployment (Nanayakkara and Hullman, 2023). Operationally, adjustment and differential privacy were also complex in that they impacted many teams and pipelines within the data making process. They not only were perceived to have large impacts on the data products themselves, but also on the operations that produced the data. In terms of adjustment, this led Census Bureau statisticians to opt not to move forward with adjustment even before the Secretary of Commerce squashed the effort. On the other hand, differential privacy moved forward even with significant hesitation around the reliability of implementing it.

Interestingly, the networks of opposition were structured differently in each case. Early on in the "adjustment wars," there was not a partisan or ideological divide among actors. Once politicians concluded that adjustment would benefit some communities more than others, the methods for adjustment became highly political, resulting in a long stretch of court cases (Choldin, 1994). Even statisticians—both at the bureau and externally—disagreed over whether this method would produce higher quality statistics. Fractures within the network increased over time, polarizing stakeholders and tearing apart the network that helped ensure the legitimacy of the census. We were repeatedly told that narratives of complexity around implementation and debates over the impact of the methods themselves made adjustment not only externally but also internally contested.

As the adjustment war unfolded, the networks of stakeholders shifted and fragmented. Unlike other government controversies where partisan cleavage was driven by the press (Bach and Swartz, 2022), here the media, courts, and other interested parties all fed off of each other to drive the partisan rearrangement.  In the end, Democrats and civil rights advocates were in favor of statistical adjustment while Republicans opposed it; data users and statisticians were split. This helped trigger closure; discussions to re-open statistical adjustment are immediately viewed as partisan and non-negotiable.

With differential privacy, the arrangement of stakeholders was different. Partisan actors were generally neutral to opposed while attitudes among data users, statisticians, and civil rights advocates varied widely. As of this writing, differential privacy is not viewed along partisan lines even though the first two lawsuits had a partisan orientation.  There is also no consistent or logical clustering of stakeholders by role or ideology.

One important difference in the network formations between the battles over adjustment and differential privacy is with regard to the time-scale of implementation. Adjustment was debated and contested over decades, eventually failing to be deployed, which created enough time for the network to split along known lines in response to this method (Anderson and Fienberg, 1999). Differential privacy, on the other hand, was announced and deployed within a span of a few years, which many opponents argued did not enable them to it to block its deployment in time.

While many stakeholders felt that differential privacy should have been introduced over a longer timeline, in order to create more consensus around its implementation (Hotz and Salvo, 2022), it is likely that such a ramp-up may have allowed polarization to take root. Of course, given the long timespans between each census, the network may be able to reconfigure itself towards a different outcome in 2030.

To some statisticians, adjustment was a natural extension of imputation, and differential privacy was an inevitable modernization of swapping. This is indicative of a view that census methods are and should be continually evolving alongside statistical knowledge. For those who see the work of a census as an accounting project, however, this orientation towards statistical advancements is less natural. Herein we can see cleavages in the statistical imaginaries that orient stakeholders. This became especially pronounced as stakeholders navigated highly visible and highly complex statistical approaches. What emerged set the scene for fierce contestations that—in some cases—spilled over beyond these two methods.

**5.3** *Imputation and swapping: shifting dynamics and reconfigured arrangements*

While swapping and imputation were not initially controversial, political forces shifted attitudes towards these methods later, during controversies about adjustment and differential privacy—two methods that were more rhetorically and operationally complex, and whose visibility was much greater. In particular, imputation was scrutinized and brought to court only after it became known through adjustment debates, while swapping flew under the notice of most data users until the Census Bureau decided to move to differential privacy. What made imputation and swapping controversial at that point was that the networks and rhetorical claims became destabilized by these *other* techniques.

As the need to deal with missing data and disclosure avoidance grew, mathematical statisticians started advocating for different methods. In their eyes, if the goal was to correct undercounts, then population-level adjustment might provide better accuracy across different levels of geography than does hot-deck imputation. And if the goal is to protect the confidentiality of data subjects while providing accurate data, then using principled methods for noise infusion—such as those that satisfy differential privacy—do so better than swapping. Complexity of method did not intimidate the experts at the Census Bureau. Yet, pressure to shift the methods increased the visibility of the accepted methods while also taking for granted the legitimacy conferred by decades of use.

It is impossible to know if imputation and swapping would have become controversial if not for adjustment and differential privacy. Once these latter practices shone a spotlight on the former methods, however, they exposed gaps between rhetoric and practice. Despite stakeholders' perception that imputation was used at a small scale, the application of this method has been steadily increasing. Due to ballooning non-response rates, imputation was used at a larger rate than ever in 2020 to fill in missing records. Similarly, as the threats to privacy have escalated over the last twenty years, the rate of swapping would have needed to scale up accordingly. While the Census Bureau did not apply swapping to the 2020 data, it said that high swap rates—even up to 50%—would still only have a minimal protection against re-identification attacks while reducing accuracy significantly (Hawes and Rodriguez, 2021). In other words, continuing to use swapping would have a "significant, detrimental impact on data quality" (Abowd and Hawes, 2023).

Here is where our two cases diverge. After adjustment was litigated and the Census Bureau opted to not pursue this method, states turned to litigate imputation, culminating in Utah v. Evans (2002). The Supreme Court's decision preserved the use of hot-deck imputation, but prevented future adoption of model-based imputation techniques that would result in higher data quality.

The story of swapping unfolded differently, in no small part because litigation (Alabama v. United States Department of Commerce, 2021) did not prevent differential privacy from being implemented. As a result, stakeholders who abhorred differential privacy doubled down on the value and importance of swapping, refusing to engage with whether it was viable at scale even as researchers highlighted that pursuing this method would have significant implications on accuracy (Christ et al., 2022). In other words, the networks and rhetoric around swapping remained sticky even as differential privacy was deployed.

## 6. From contested methods to legitimate futures

The analysis we have laid out teases out the various arrangements of statistical imaginaries in each case, showing that arrangements—not singular imaginaries—are critical for understanding not only the reception to methodological interventions, but also how modernization efforts impact trust in government data infrastructures. The entanglements between networks and rhetoric, visibility and complexity, help us analyze why certain statistical methods are adopted or foreclosed, how arrangements of imaginaries are reinforced or reconfigured, and what this means for the legitimacy of critical data projects like the census.

Each of the four methodological advances we traced has had different receptions. After decades of contestation, the Census Bureau's adjustment program was foreclosed. Once adjustment was shut down, the controversy spilled over onto imputation. As a result of the lawsuits that followed, the techniques for imputation became frozen in time. And with the modernization of disclosure avoidance, swapping was replaced entirely. But as of this writing, the story is not over around differential privacy or even imputation. Litigation threats continue to hover in the background.

Differential privacy was a flashpoint of the 2020 census, but it is also a success story in deploying a complex method in an environment of extreme distrust and politicization. At the same time, the high visibility of this complex method shattered statistical illusions around census data that were critical to both the network structure and the rhetorical claims of the existing arrangement. The arrangement of statistical imaginaries underpinning the legitimacy of the census is still fractured. It is yet to be seen whether contestations around differential privacy will remain a chapter of the past, whether its implementations will cease to be updated as with imputation, or whether like adjustment the fights around disclosure avoidance will become central in the politics of future censuses. It is also unclear if, when, or how the arrangement of statistical imaginaries will become stable again.

What we do know, more clearly than ever, is that what legitimizes the work of the Census Bureau is more than its production of "the best possible count" of the population. Each decadal count is made credible by myriad practices and illusions, many of which are derived from and serve to stabilize the unique, and increasingly precarious, arrangement of statistical imaginaries. Computer scientists advocating for differential privacy in the 2020 census learned what sociotechnical scholars already understand: that to privilege technical methods over socio-political factors is to misunderstand the entire project of the census. Those who wish to improve the count must not only envision how to make "better" data, but also how to knit together the diverse perspectives, agendas, and politics around the census.

Our analysis builds on the long literatures on quantification and the state by accounting for how arrangements of statistical imaginaries shape and are shaped by controversies that appear, at first blush, to be about statistical methods (Bouk and boyd, 2021). These literatures have long recognized how durability of collective visions impacts the credibility of state numbers, and vice versa. Our work goes further to investigate the building blocks of this critical, recursive relationship. We highlight how *multiple* entangled imaginaries mediate the relationship between quantitative data and the state. Looking at that the networks, rhetoric, and *visibility* and *complexity* of each method can help us analyze how well the statistical imaginaries of the census "hold together" (Desrosières, 1990).

When new statistical methods are introduced, their reception depends heavily on how disruptive they are to the arrangement of imaginaries. This depends on the rhetoric that forms around them and the ways in which networks reposition themselves in response. In addition, those advocating for new statistical methods should think deeply about how easily these methods can be leveraged to disrupt the arrangement. When new methods are highly visible and complex, much more work is needed upfront to stabilize networks and gain rhetorical buy-in towards collective visions. At the same time, calcifying statistical practice to appease critics will not resolve statisticians' commitment to modernizing statistical projects as the science evolves. When statistical practices are kept hidden or confined to outdated assumptions about the data, they entrench brittle arrangements of imaginaries and set the stage for outsized battles over modernizations in the future. Thus, our analysis acknowledges the need for transparent modernizations even while we caution actors to recognize the fragility of arrangements—best understood by longstanding census data users—and invest in the work required to build collective visions. Without shared orientations, new statistical practices—and politically motivated actors who can trigger and leverage shifts in arrangements—will only serve to weaken the legitimacy of the project.

In the short term, the goal is to smooth divides by engaging with, listening to, and learning from those on opposing sides. In the long term, the goal is to bring together different communities around the census to build more robust, flexible arrangements of imaginaries around the census that will be receptive to methodological advancements. Without such repairs, new conflicts that are similar to these old conflicts will continue to emerge. Political efforts will ensure that such efforts delegitimize the statistical work and solidify antiquated methods into place.

The work of building trusted data infrastructures is challenging, especially when the data is designed to sit at the heart of a nation's social and political life. The U.S. Census Bureau may always need to rely on statistical illusions of objectivity and precision in order to exist. But for those who want the census to not just provide data that is trusted, but also *trustworthy*, we must grapple with *how* and *why* methodological advances—from imputation to adjustment, swapping to differential privacy—rupture or rebuild legitimacy around the census. Understanding the complex orientations towards data can help us create not only better numbers, but also more stable arrangements to support them.

**Acknowledgements**

This work was made possible because of the countless civil servants at the Census Bureau and census stakeholders who took the time to help us understand the different techniques the bureau used and the politics that shaped them. We are also grateful to Kevin Ackermann and Parker Bach for their valuable contributions as research assistants on this project. This work was enriched by feedback from participants in the 2022 Privacy Law Scholars Conference, 2022-23 Edmond & Lily Safra Center for Ethics seminar, 2023 ASA Session on the Interactions of Data & Society, and countless colleagues including Dan Bouk, Julie Cohen, Catherine Crump, Alex Hanna, Chuncheng Liu, Priyanka Nanayakkara, Ryan Steed, Janet Vertesi, and Moira Weigel. The first author was funded by a Columbia Data Science Institute fellowship. The second author was funded by her home institution, with additional support from the Alfred P. Sloan Foundation, the John S. and James L. Knight Foundation, and Data & Society.